\documentclass[aps,twocolumn,showpacs,preprintnumbers,amsmath,amssymb,floatfix]{revtex4}
\usepackage{graphicx}
\usepackage{epsfig}
\begin{document}
\def\be{\begin{equation}}
\def\ee{\end{equation}}
\def\bearr{\begin{eqnarray}}
\def\eearr{\end{eqnarray}}
\def\half{$\rm \frac{1}{2}$~}

\title{Singlet Reservoir Theory of Ambient Tc Granular Superconductivity in\\
Monovalent Metal Nanostructures}

\author{G. Baskaran}
\affiliation
{The Institute of Mathematical Sciences, C.I.T. Campus, Chennai 600 113, India \&\\
Perimeter Institute for Theoretical Physics, Waterloo, ON N2L 2Y5, Canada}

\begin{abstract}
Monovalent metals contain half filled band (HFB) of s-electrons. Emphasizing importance of Coulomb repulsions in HFB in 2D and 1D monovalent systems we sketched a theory (2018) for ambient temperature granular superconductivity reported by Thapa and Pandey (2018) in Au-Ag nanostructures (updated by Thapa et al., 2019). Sharpening our theory, we suggest that \textit{Coulomb repulsions in certain structurally perturbed regions (atomic clusters, stacking faults, grain boundaries etc.) create nanoscale reservoirs of singlet electron pairs}. These low dimensional patches are hybridized to a background 3D jellium metal and produce observed ambient Tc granular superconductivity via proximity Josephson effect. Using repulsive Hubbard model we show presence of singlet reservoirs and physics of doped Mott insulators. Needed charge transfer arises from differing electronegativities. Our theory predicts that \textit{all elemental monovalent (alkali, Cu, Ag and Au) metals, under suitable structural perturbations, are likely to exhibit ambient temperature superconductivity}. 
\end{abstract}
 \maketitle

\section*{I Introduction}
 
Superconductivity, a remarkable macroscopic quantum phenomenon, touches unexpected and different corners of physics and science in general. Its technological implications are also profound. Discovery of such a phenomenon in Hg in 1910, by Kamerlingh Onnes and decades of developments that followed have nurtured a strong desire and dream to realize superconductivity at ambient temperatures and ambient conditions.

A milestone discovery by Bednorz and Muller \cite{BednorzMuller1986} of high Tc superconductivity in cuprates, in 1986, suggested that this dream could be realized. Resonating valence bond theory of high Tc superconductivity that began as a response to understand cuprate superconductors pioneered by Anderson \cite{PWA1987,BZAandOthers,KRS,PlainVanila,GBIran} gave new hope and created a fertile ground for high Tc superconductivity, quantum spin liquids \cite{WenBook,TaiKaiSpinLiquid} etc. Mott insulators, as a platform for action, played a key role. On the experimental front new families of high Tc superconductors have been discovered. Ambient Tc's have been reached in pressurized Lanthanum hydride recently \cite{Somayazulu,Eremets,GBH2S}.

In this background a report of signals for ambient temperature granular superconductivity \cite{ThapaPandey1} in Ag nanoparticles embedded in Au matrix came as a surprise last year (2018). After a period of silence Thapa and collaborators at IISc (Bangalore, India) have presented a new report \citep{ThapaPandey2} claiming validation of earlier findings in Ag-Au nanostructures (Ag-Au NS)

This discovery is puzzling, given the well known fact that all non-transition monovalent metals resist superconductivity, even at the lowest temperatures (Li is an exception: Tc $\approx$ 0.4 mK). A large Coulomb pseudo potential \cite{PWAMorel} prevents occurrence of phonon mediated superconductivity. Three dimensional monovalent metals are robust Fermi liquid metals; electron correlation effects produce vanishingly small Tc \cite{KohnLuttinger,RaghuSteveDough,MaitiChubukov}.

We present a theory, a sharpened version of our earlier work \cite{GBv2} for the observed exciting phenomena in Ag-Au NS. Ours is a semi quantitative theory that focusses a matter of principle - Is ambient Tc superconductivity possible at all, within known quantum condensed matter physics in Ag-Au NS system ? We focus on what we suggest to be a natural electronic mechanism and proceed in four steps:

i) a structural model suggesting presence of quasi localized interacting electron subsystems present as nanoscale 2D patches, 1D segments and 0D clusters coupled to a jellium metallic matrix is presented, ii) we show that, because of monovalent character, they become a Mott-Hubbard network of 2D, 1D and 0D nanoscale singlet reservoirs hybridized to jellium metallic matrix, iii) emergence of strong local superconducting pair correlations, via charge transfer and doped Mott insulator physics and iv) a model of proximity effect induced Josephson coupled network.

Our model focusses on certain unavoidable structural modifications present, at and away from interfaces of Ag nano particles embedded in Au matrix. They are, stacking faults, dislocations, grain boundaries, embedded clusters etc. Preparation method, differing quantum chemistry of Ag and Au, their diffusion, metallurgical intricacies etc., contribute to this. 

Monovalent metals are effectively single band tight binding half filled band system, created by monovalent s-orbital overlap \cite{NaMagnet,Goddard,Vanderbilt,AuAgClusterBulk,WFCuAgAu}. They have been studied using single band repulsive Hubbard model. However, Hubbard repulsion U is inconsequential for 3D bulk Ag and Au and they get renormalized away at low energy scales, leaving behind a robust Fermi liquid. Situation is comfortingly different in 2D,1D and 0D repulsive Hubbard models close to half filling. Fermi liquid gives way to doped Mott-Hubbard physics and generates strong local superconducting pair correlations.

In our theory, necessary internal charge transfer (self doping) arises from differing electronegativities. Proximity effect via embedding metallic matrix is modelled as a Josephson network of granular superconductors. We also use a theoretical model of Berg, Orgad and Kivelson \cite{BOK} (Route to high-temperature superconductivity in composite systems) in our theory for monovalent metal system, to estimating Tc's. A work by Chen et al. \cite{CSTing} on proximity is also used: this work discusses possibility of enhancement of superconducting Tc in a strongly correlated tJ layer, when it is hybridized to a free electron layer, under some conditions.

It also follows from our theory that \textit{elemental monovalent metals (alkali, Cu, Ag and Au), under suitable structural perturbations are likely to exhibit ambient Tc granular superconductivity}.

In the present paper we focuss on doped Mott insulator physics that has proven validity in the context of single band system such as cuprates and other strongly correlated superconductors, consistent with our earlier proposal of `Five fold way to new superconductors' \cite{GB5FoldWay} and emergent Mott insulators in effectively monovalent systems \cite{GBSilicene,GBH2S}

Mechanisms such as phonon exchange and plasmon exchang, in our estimate, does not create enough pairing strength to produce ambient temperature pairing in our system containing strong structural perturbation. It is also important to note that Au under certain structurally perturbed conditions exhibits Ferromagnetism \cite{AuFM} - it underscores importance of electron-electron effects in structurally perturbed Au and monovalent systems.

Our paper is organized as follows. After a brief summary of recent experimental observations we present arguments for existence of nanoscale 2D patches, 1D segments and 0D clusters in this otherwise 3D nano composite structure. Then we briefly discuss how superconducting correlations arise in these nanoscale low dimensional systems using physics of repulsive Hubbard model. Having identified the sources we present a model of Josephson coupling via proximity effect from embedding metallic matrix. 

\section*{II Summary of Signals for Ambient Tc\\Granular Superconductivity in Ag-Au NS.} 

In the work by Thapa and Pandey \cite{ThapaPandey1}, and in the recent updated and refined experiemnts by Thapa and collaborators \cite{ThapaPandey2}, nanoparticles of size $\sim$ 10 nm get embedded into a Au matrix. A sharp resistivity drop to values, few orders of magnitude below values of best known metals, at ambient temperatures is seen. There is also Meissner like diamagnetism that accompanies the resistivity transition. It is not a perfect diamagnetism - this has been attributed to granular superconductivity with a small Meissner fraction. It is also interesting that Tc is sensitive to external magnetic fields. In some samples, the above Meissner like signals are seen above room temperatures, indicating a great potential for increase of Tc.

\section*{III Structure of Nanoscale 2D, 1D and 0D\\Singlet Reservoirs in Ag-Au NS}

In our theory certain nanoscale lower dimensional defects in Ag-Au NS host localized superconductivity by being reservoirs of singlet electron pairs. We present a structural model for lower dimensional defects for Ag-Au nanostructure which are potential singlet reservoirs.

Structural modifications in Ag nanoparticle embedded Au matrix arise from preparation methods and differences in solid state, quantum chemistry of Ag and Au and metallurgical complexities. From available structural information \cite{ThapaPandey2} one can infer to some extent nature of defect content inside Ag nanocrystal we are interested in. 

In the supplement section we identify relevant structural modifications which are potential singlet reservoirs. HRTEM image (figure 1) \cite{ThapaPandey2}) of a single Ag nanocrystal (when magnified) shows stacked 111 planes, misoridented nanoscale patches of 111 planes, nanoclusters, stacking faults and so on . An important point to note is that size of nanoparticles provides a cut off to the size of defects we can have, as size of Ag nanoparticles and inter nanoparticle distances are in the scale of 10 nm. 

Structural reconstructions are possible in Ag and Au. Indeed reconstruction of fcc lattice, via change of stacking sequence to hcp, 4H and 9R structures and other structures are known \cite{LiHoffmann,WaghmareAyyubAg,9RNumerics,Cu9R,Ag9R}. 9R structure has been seen in experiment \cite{Ag9R} for Ag and Cu. Further, band structure studies of 9R structure for Li \cite{LiHoffmann} show quasi 2D Fermi surfaces. 

Chakraborty and collaborators \cite{WaghmareAyyubAg} have studied 4H form of Ag experimentally and theoretically. 4H, with higher resistivity (compared to fcc) also has a 10 fold, c-axis to ab-plane, resistivity anisotropy. What is interesing is occurrence of 2D Fermi surfaces in the BZ. An emergent local 2d Fermi surface close to half filling could support RVB type superconductivity.

In supplementary section II, we discuss a model stacking (ABA repeat) and illustrate how effective electronic decoupling of layers leading to 2 dimensionality emerge in 4H and 9R structures.

In the Ag-Au composite one expects, a charge transfer from Ag to Au, close to the interface of nano crystal and bulk matrix. Electronegativity (ability to attract an electron) is higher for Au (2.54) than Ag (1.93). So we expect electron transfer from Ag to Au. Consequently, Ag will be hole doped and Au, electron doped. Metallic screening is likely to keep doping close to grain boundaries and Ag-Au interface regions. Because of structural and bonding changes near structural perturbations, local transfer is possible within Ag and Au regions. 

It is also known that Ag and Au (111) surfaces support 2D surface bands, which are far from half filling, with a Fermi energy of 0.5 eV and band width about 4 eV. If there is sufficient charge transfer it could support strong pairing correlations.

Given the fact that size of Ag nanoparticle is $\sim$10 nm and inter distance $\sim$ 15 to 20 nm, in our estimate, a liberal density of 2D, 1D and 0D defects are available to give a rich granular superconductivity physics. We will see in the next section how required strong local superconducting pairing arises in Ag-Au NS.

\section*{Estimation of Pairing Energy Scale in\\Singlet Reservoirs}

A single band repulsive Hubbard model is suitable for 3D monovalent metals, even though Hubbard U becomes irrelevant in the sense of renormalization group theory. At low energy scales we have a robust Fermi liquid, a jellium metal with nearly spherical Fermi surfaces. 

Situation is different in 2D, 1D and 0D clusters, where Coulomb repulsion brings in new physics. We get Mott localization in 1D for arbitrarily small repulsion. In 2D we get enhanced pairing correlations close to half filling for intermediate U, even below the Mott insulator region (figure 1).

It should be pointed out that there is no rigorous proof for existence of superconductivity in repulsive Hubbard model in 2D, in the Mott insulator regime, close to half filling. Hoever, physics motivated approaches such as RVB theory and manybody theory techniques point to possibility of strong local singlet electron pairing. They provide good estimates for pairing temperature scales, as a function of Hubbard model parameters.

In what follows we take model parameters for Ag and Au from the work of Torrini asnd Zanazzi \cite{TorriniZanazzi} on monovalent clusters. Nearest neighbor hoping t $\approx$ 1,9, 2.3 eV for Ag and Au and onsite repulsion U $\approx$ 6.75 for both. 

\textbf{Atom Clusters as Singlet Reservoirs:} Monovalent metal atom clusters, including Ag and Au have been extensively studied from the point of view of electron correlation effects and magnetism \cite{TorriniZanazzi,Muhlschlegel,Pastor,NobleClusters,deHeerClusterRMP,PuruJena,HudaRay}. Strong antiferromagnetic exchange J $\sim$ 1.67 and 2.33 eV (for Ag and Au) gives rise to strong singlet bonds and resonances. In comparison to cuprates, where J $\approx$ 0.15 eV, we have an order of magnitude larger pairing energy scale. Consequently local pairing temperature scale are several hundreds of Kelvin.

Effective Hamiltonian for a monovalent atom cluster is the Heisenberg antiferromagnetic Hamiltonian:
\be
H_c = \sum_{ij} J_{ij} ({\bf S}_i\cdot{\bf S}_j - 1/4) \equiv -\sum_{ij} J_{ij} 
b^\dagger_{ij}b^{}_{ij} 
\ee

Spin operator ${\bf S}_i \equiv d^{\dagger}_{i\alpha} {\hat\sigma}_{\alpha,\beta}d^{}_{j\beta}$.
Spin-spin interaction term has been rewritten using bond singlet operator
$b^\dagger_{ij} \equiv (1/\sqrt{2} (d^{\dagger}_{i\uparrow}d^{\dagger}_{j\downarrow} - d^{\dagger}_{i\downarrow}d^{\dagger}_{j\uparrow})$, (d's are electron operators) to emphasize the fact that strongly coupled bond singlets are natural basic units (rather than spin moments) in clusters. It is the resonance of these bond singlets which is at the heart of our mechanism for superconductivity in Ag-Au NS. More importantly resonance energies of bond singlets are comparable to cohesion energy of monovalent clusters \cite{Pastor}.

Consider a simple tetrahedral cluster, an elementary closely packed unit in fcc lattice. Two singlets resonate and stabilize a singlet ground state. Larger clusters have been studied from the point of view of equilibrium shapes of clusters. For us what is important is the singlet content and their resonance. The tetrahedral RVB system has a doubly degenerate singlet ground state.

We have suggested presence of a distribution of clusters in Ag-Au NS. As we go to larger clusters there are more complex resonances, depending on size and shape of a cluster. Singlets and their resonances continue to be present till a critical size. Beyond the critical size singlets and their resonances go off the shell \cite{GBv2} and the bulk of the cluster behaves as a jellium metal (filled shells) at low energies. Surfaces may continue to have interesting singlet correlations and resonances.

It is clear that clusters of various kinds, in the Ag rich, Ag rich regions and in the interfaces are very important as singlet reservoirs. We also have a very large singlet pairing energy scale J $\approx$ 1 eV.

\textbf{Chains as Singlet Reservoirs}
Consider isolated Ag or Au chain \cite{AtomChainGoddard,MetalChainBook}. If we ignore lattice dimerazation instabilities, 1D Hubbard model predicts a Mott insulator with a Mott Hubbard gap of $\sim$ 1 eV (Lieb and Wu \cite{LiebWu}) and superexchange J $\sim$ 1.67 and 2.33 eV for Ag and Au. Effective Hamiltonian of a Ag or Au chain is given by an antiferromagnetic Heisenberg Hamiltonian similar to equation 1.

Resonating singlets in the spin-\half chain become our reservoir singlets. An organization of
Ag or Au chains into weakly interacting chain and doping easily brings in ambient Tc superconductivity as discussed below.

We consider a 2D array of weakly coupled (for a range of interchain hopping t$_\perp$) Ag or Au chains. To avoid competing antiferromagnetic 2 sublattice order, we stagger the chains,so that when t$_\perp$ = t we get isotropic triangular lattice. Using known studies  we find doped Mott insulator physics and ambient Tc superconductivity, provided competing instabilities such as dimerization and valence bond order are taken care of. 

In the above system pair tunneling plays an important role in stabilizing 2D or 3D superconducting order. We have used Wheatley-Hsu-Anderson's mechanism \cite{WHA} of interlayer pair tunneling and similar interchain pair tunneling theory of Zachar, Kivelson and Emery \cite{ZKEmery} to estimate superconducting Tc's.

Scale of Tc is given by k$_{\rm B}$Tc $\approx$ t$^2_\perp$/$\Delta_s$. Here $\Delta_s$ is a local spin gap scale. It can be approximated by superexchange J, in the present context. 
It is easily seen that for reasonable choice of t$_\perp$, in our model system of Ag and Au we get 2D superconducting Tc at ambient temperature scales. Thus we conclude that nanoscale line segments in Ag-Au NS are are also important singlet reservoirs, with a high pairing temperature scale.

{\bf 2D Plane Segments as Singlet Reservoirs:}
Two dimensions is equally interesting. We considered hypothetical 2D triangular lattices of Ag and Au. Firstly, U is not strong enough to result in Mott localization. However, a form of (self) doped Mott insulator physics (dynamical Mott localization, figure 1)  survives at and close to half filling. Using results from existing body of literature we estimated Tc  \cite{TcThomale,TriangularTc1,TriangularTcTremblay,TriangularTcOhta,SumitMazumdar,Tanmoy,TcJarrel} 
. After doping, results for Tc are not very different for triangular lattice and square lattice, even though symmetry of order parameters are different (d-wave for square lattice and d + id for triangular lattice.

Interestingly, in these model studies k$_B$Tc is a finite fraction of hoping parameter t, even for intermediate repulsions. In view of this non-BCS dependence we obtain ambient temperature (and higher) pairing temperature scales for Ag and Au subsystem at and close to half filling.  We emphasize that in real  2D systems, when synthesized, competing orders such as valence bond order and lattice dimerization are big hurdles to reach ambient Tc superconductivity.

A schematic phase diagram of repulsive Hubbard model in 2D triangular lattice is shown in figure 1. We name the region close to half filling, below U$_c$ as dynamical Mott localization, as it resembles a self doped Mott insulator in its properties, such as existence of electron correlation based superconductivity in the absence of nesting induced spin fluctuations and correlations.

\begin{figure}%[htp]
\includegraphics[width=0.6\textwidth]{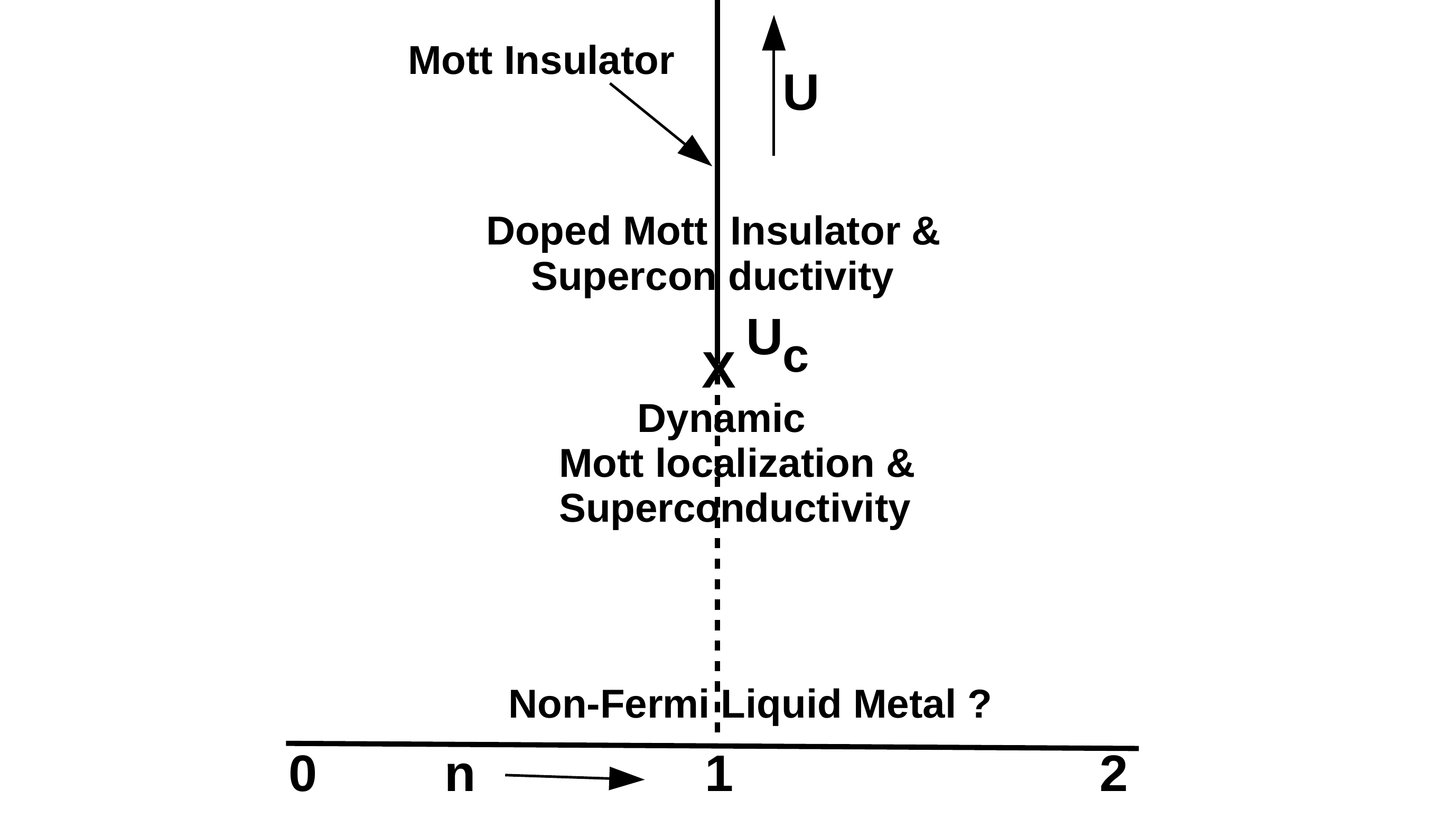}
\caption{Schematic phase diagram for repulsive Hubbard model on a 2D triangular lattice, as a function of U and site occupancy n. Region below
U$_c$ has quantum melted (dynamical) Mott localization or what has been called
\textit{self doped Mott insulator} \cite{GBOrganics}. According to Anderson \cite{PWATLLiquid} Fermi liquid theory fails in 2D even for small U resulting in a Tomographic Luttinger Liquid.}\label{Figure 1}
\end{figure}

What is important is that the nanoscale 2D patches continue to support strong singlet pairing,
which could eventually contribute to observed 3D ambient Tc granular superconductivity.

\section*{Stability of Reservoir Singlets}
An important question is survival of reservoir singlets, when there is hybridization to the 3D embedding metallic matrix. To gain some insights into this issue we introduce a simple model and discuss results qualitatively. Consider a tetrahedral cluster coupled to a 3D metallic matrix. It is a tJ model of tetrahedral cluster coupled to 3D jellium metal: 

\bearr 
H &=& \sum_{k, \sigma} (\epsilon_k - \mu) c^\dagger_{k\sigma}c^{}_{k\sigma} +\nonumber\\
&-&t \sum_{\sigma, i,j = 1}^4(d^{\dagger}_{i\sigma}d^{}_{j\sigma} + H.c.) - J \sum_{i,j = 1}^4 b^\dagger_{ij}b^{}_{ij} +
\nonumber \\
&+& \sum_{\sigma,k,i = 1}^4(f_{k}(i) c^\dagger_{k\sigma}d^{}_{i\sigma} + H.c.)
\eearr 
Here c's and d's represent operators of metallic plane wave states and operators of four atomic sites of our tetrahedral cluster.  Hybridization f(k,i), hopping t and kinetic exchange J have usual meaning. We also have appropriate local constraints on occupancies at sites i. 

In an isolated tetrahedral cluster, number of valence electrons is 4 and fixed. We have a doubly degenerate ground state with resonating singlets (for suitable linear combinations they become two degenerate and PT violating chiral singlets). Once we allow hybridization through the 3rd term in the above equation we have a possibility of charge fluctuation and hopping t within the tetrahedral cluster, in addition to the exchachange J. Depending on position of atomic level with respect to chemical potential of the metal, electrons or holes are allowed to enter the tetrahedral cluster and hop around. We control \textit{doping} via local occupancy constrains on occupancies of 4 atoms and chemical potential which determines mean number of electrons in a cluster.

We have some general conclusions for this model: i) Singlet correlations within the tetrahedron decreases with increase of hybridization and ii) singlet pairing correlations will spread to metallic matrix. 

We have also looked at model, a 1D Mott-Hubbard chain weakly hybridized to planewave states of 3D metal. As anticipated on physical grounds, stability of singlet reservoir and Mott localization survive till a critical hybridization strength. Interestingly, singlet correlations survives as dynamical Mott localization on the metallic side, close to metal insulator transition. We hope to study this model in detail in the future. 

Fortunately we can answer our question of stability of reservoir singlets using certain known experimental result as well. Clark and collaborators found \cite{4Molecules} the following remarkable result in the context of organic superconductors. They deposited a number of organic molecules, a 2D lattice patch, on Ag (111) substrate. This is a finite size system of a well known organic superconductor, where superconductivity is based on electron correlation mechanism (described by 2D repulsive Hubbard model). Using STM measurements they studied superconducting gap as a function of number of molecules within a patch. They found surviving  superconducting gap like structure till number of molecules became as small as four !

This experimental result demonstrates that a spin singlet electron pair reservoir, as small as 4 molecules, remains statble in the presence of coupling to a metallic substrate. 

\section*{Proximity Effect, Web of Reservoir Singlets \&\\ Ambient Tc Granular Superconductivity}

Two macroscopic superconductors connected by a weak link exhibit Josephson effects, via Cooper pair tunneling. Weak link is typically an insulating barrier. In our case we have nanoscale patches (which support pairing correlations and not real superconductivity) embedded in a 3D metal matrix. It is known that such type of system can have Josephson like coupling through proximity effect and produce macroscopic superconductivity. 

we present a physical picture for proximity effect in nanoscopic and phase fluctuating singlet paired system.  Consider two electron wave packets with opposite momenta with random spin orientations from the jellium metal enter a nanoscale patch simultaneously. As these electrons are identical to interacting electrons inside the doped Mott insulator patch, they will experience strong singlet pairing and come out dominantly in spin singlet channel. Thus singlet reservoirs act like singlet filters and and also transfer and share their spin singlet correlations to the jellium metallic matrix in a dynamical fashion. 

Proximity effect in monovalent metals like Ag and Au is also known to have unusually long length scale \cite{ProximityAg}. Consequently reservoir singlets have effective Josephson coupling which is long ranged. Reservoir singlets connected in complex long range fashion creates for us a \textit{web of Reservoir singlets.}

It is also clear that relative volume occupied by clusters, 1D and 2D nanoscale patches in bulk Ag-Au NS is very important. A measure of this  the density of singlet reservoirs.

We formalize our problem using the following model. A free electron system representing the jellium metallic matrix is hybridized to a doped t-J model subsystem representing our low dimensional correlated electron subsystem.
\bearr 
H &=& \sum_{k, \sigma} (\epsilon_k - \mu) c^\dagger_{k\sigma}c^{}_{k\sigma} +\nonumber\\
&-& \sum_{i,j \sigma} t_{ij} (d^{\dagger}_{i\sigma}d^{}_{j\sigma} + H.c.) - \sum_{i,j}J_{ij}
b^\dagger_{ij}b^{}_{ij} +
\nonumber \\
&+& \sum_{i,k,\sigma}(f_{k}(i) c^\dagger_{k\sigma}d^{}_{i\sigma} + H.c.)
\eearr 
As in equation 3, hybridization f(k,i), hopping t$_{ij}$ and kinetic exchange J$_{ij}$ have usual meaning. In the presenct case they are spatially random. We also have appropriate local constraints on occupancies at sites i.

A quantitative study of the model we have presented above can be performed using various approximations. In an earlier work Berg, Orgad and Kivelson \cite{BOK} (BOK), studied proximity effect mediated superconductivity in a metallic matrix containing islands supporting pairing correlations. BOK were inspired by several anomalies related to superconducting pairing and Tc in underdoped cuprate superconductors, a more complex oxide system. This work, which also motivated a route to high Tc superconductivity, is eminently suited to anlyse our present situation. 

In BOK model pairing within islands is replaced by an effective attractive U Hubbard model. An important result from BOK is that \textit{bulk granular Tc can be as high as the mean field Tc present within a given patch}.

As we mentioned earlier our mean field pairing Tc is above ambient temperatures, for given t and U for our nanoscale low dimensional susbsystems in Ag-Au NS. Using the bounds on Tc from the work of BOK we conclude ambient Tc granular superconductivity is a possibility, within our model and theory.

\section*{Can 3D Metal Matrix use Proximity Effect \&\\ Boost Tc Beyond Mean field Tc of Patches ?}
In a very interesting work Chen et al. \cite{CSTing}, studied proximity effect induced superconductivity in the following model, using Gutzwiller approximation methods. Their tight binding model had a bilayer of square lattices and an interplane hopping. One layer is a free electron layer. The other is a strongly correlated system described by a t-J model, a dopped Mott insulator.

They found a remarkable result that under certain conditions a synergy from free electron system, via proximity effect can even increase the Tc of the t-J layer. We interpret their result as follows. If the free electron system has sufficiently high density of states it can take advantage of RVB pairing offered by the t-J layer and have a Tc larger than the t-J layer.

From the work of Chen et al., we conclude in our Ag-Au NS the free electron or jellium 3D background could play important role in bringing Tc of the granular superconductor to large scales, comparable to the large pairing temperature scale we have found to be present in the nano scale low dimensional singlet reservoirs.

\section*{Structurally Perturbed Monovalent Metals \&\\Ambient Temperature Superconductivity} 
A closer look at our theoretical study reveals that two important criteria to reach ambient temperatuture superconductivity are: i) monovalent character of Ag and Au and ii) low dimensional nanoscale electronic subsystems we can create, which are potential singlet reservoirs.

Thus our theory predicts that \textbf{all elemental monovalent (alkali, Cu, Ag and Au) metals, under suitable structural perturbations, are likely exhibit ambient Tc superconductivity}. Sharp structural perturbations seem to be necessary. Monovalent Alloys with sharp (atomic scale) compositional superstructures might help.
 
We plan to discuss our provocative prediction in some detail in a future pubication.

In this context a straightforward prediction of our theory is that a single atom thick monovalent metal layer (a flattened atom thick monovalent metal) is a potential ambient temperature superconductor, if competing orders such as valence bond order and gap opening lattice instabilities can be taken care of. 

We wish to refer to several diamagneitic anomalies known in alkali metals, Cu, Ag and Au \cite{AlkaliDia,CuDiaMSRao,ImryDiamagnetism,AuDiaNanorod,AuDiaNanorodRhee,AgCuCompositeDia,AgRingDia}
Signals of diamagnetism, seen in Au and Cu nano rods etc., in the past decade, at ambient temperatures are exciting reports, from the point of view of our prediction. 

\section*{V. Discussion}

Inspired by the recent discovery of strong signals for ambient temrtiperature  granular superconductivity in Ag nanoparticles embedded in Au matrix, we have presented a theoretical analysis in support of their discovery. Our main thesis is that a robust Fermi liquid in 3D, in monovalent metal becomes a metal containing dynamical Mott localization in reduced dimensions such as quasi 2D and quasi 1D.

We have suggested few possibilities for presence of certain unavoidable reduced 2D, 1D and 0D nanoscale segments and patches in Ag-Au NS. Electron correlation effects, known to be present in 2D and 1D close to half filling is used to suggest presence of strong local pairing correlations, via RVB or doped Mott insulator physics. That is, there are lower dimensional singlet reservoirs present randomly at nanoscale in the Ag-Au NS

The 3D metallic matrix, under suitable conditions will allow strong correlation effects such as strong pairing, to survive inspite of hybridization to 3D metallic states. The intervening metal, via proximity effect couples two nanoscale patches via a Josephson type of coupling. In our theory, such coupled 3D network is responsible for the observed granular superconductivity at ambient temperatures.

In the present article our focuss is on showing theoretically a remarkable possibility of strong correlation based ambient temperature superconductivity in structurally perturbed monovalent metals. There are other important issues such as order parameter symmetry, complex superconductivity phenomenology, vortex tangles etc.

Our proposal also suggests that other nanostructures involving two or more monovalent metals could give rise to ambient Tc superconductivity. It also follows from our theory that \textit{all elemental (alkali, Cu, Ag and Au) monovalent metals, when structurally perturbed will exhibit ambient Tc granular superconductivity to varying degrees}.

To confirm our theory and predictions it will be very interesting to study isolated nanocrystals, quantum dots, rods, heterostructufes of monovalent metals, using STM and other probes, after creating controlled structural perturbations. Study of charge 2e noise will also be valuable. 

Does ambient temperature granular superconductivity exist already, in some form or other, in the rich world of minerology and metallurgy - natural ones and man made ? An exploration into minerology and metallurgy is likely to offer surprises. 

In a recent article Kopelevich, da Silva and Campargo \cite{Kopelevich} have summarised earlier appearances of ambient Tc superconductivity in experiments, under a title `Unstable and elusive superconductors'. It starts from early work of Ogg on (alkali) metal-ammonia solutions to a body of recent works on graphite \cite{Esquinazi}. A collective effort to understand these elusive superconductors, including the Ag-Au NS \cite{Awana} and make them stable with desired characteristics at ambient temperatures and conditions, will be valuable . 

\section*{Acknowledgement}

I thank R. Ganesh and M.S. Laad for discussions. A Distinguished Fellowship from Science and Engineering Research Board (SERB, India) is acknowledged. It is a pleasure to thank Perimeter Institute for Theoretical Physics (PI), Waterloo, Canada for a DVRC and hospitality. PI is supported by the Government of Canada through Industry Canada and by the Province of Ontario through the Ministry of Research and Innovation.

\section*{Note} As we were finishing this manuscript, a report came from Hooda et al., (IIT Mandi, India). They have reported \cite{Hooda1}, vanishing resistance at ambient temperatures in Ag-Au thin film nanostructures. This study in a different nanostructure, containing strong nanoscale structural perturbations, is encouraging from our theory point of view.

\section{Supplement}

\section*{I Defects Which are Potential Singlet Reservoirs} 
In addition to the Ag rich and Au rich regions we expect lots of structural defects in the interface regions between Ag nanocrystal and Au matrix. We enumerate some of the structural defects which are potential singler reservoirs.

1) \textbf{Nanoscale clusters:} Electron correlation effects are maximally revealed in nano size systems such as monovalent atomic clusters \cite{TorriniZanazzi,Muhlschlegel,Pastor,NobleClusters}. Monovalent clusters, the simplest unit, is very important as a source of electron pairing that we are after. Monovalent clusters appear inside the Ag nanoparticle, in the Au matrix and in the interface region. In the interface region monovalent clusters of Ag could be embedded in Au matrix and vice versa. In our estimate from the figure 1 in \cite{ThapaPandey2} sizes vary widely from few atom clusters to few hundred atom clusters.

2) \textbf{Grain Boundaries in Ag rich Au rich regions:} Inside Ag nanocrystals and Au matrix grain boundaries will be present. In a given Ag nanocrystal more than a dozen granin boundaries are visible (figure 1, \cite{ThapaPandey2}) as 111 planes with multiple orientations. 

Foreign atoms tend to diffuse and aggregate at grain boundaries. Close to the interface between Ag nanocrystals and Au we expect copious diffusion of Au atoms into Ag and vice versa. It will result in quasi 2D and 1D Au patches in the Ag region and vice versa. In figure S2 of reference \cite{ThapaPandey2} one does see lots of inclusions of Au atoms in Ag nanocrystals. It is not clear if it is diffusion driven or preparation method dependent.

Low angle grain boundaries are known to support arrays of dislocation lines, Dislocation lines  present in grain boundaries in Ag nanocrystals are also likely to attract lines of Au atoms, via diffusion.

3) \textbf{Decoupled ribbons}  A dislocation pair is an extra plane, of finite width, inserted between two neighboring planes in an otherwise periodic lattice. When the distance between dislocation pairs is small what we have is a ribbon of inserted two dimensional plane. Since our systems are finite, it could have an incommensurate relation with bounding planes. This could arise from small twists for example. Such planes becomes a source of 2D correlated electrons under suitable conditions. Such decoupled planes are likely to occur close to the interface. 

4) \textbf{Stacking Faults and Decoupled Planes:} Stacking faults can be formed in fcc crystals via production of dislocation pairs and separating them to the boundary. Because of low stacking fault energies for Ag and Au, a variety of stacking faults are possible. Two well known periodic stacking faults in the fcc (ABC) packing produce hcp (AB) and 9R (ABC.BCA.CAB) structures. Interestingly, a cylindrical Fermi surface emerges in tight binding model for hcp, 4H and 9r structures.

In the next section we will see how effectiv two dimensionalitys arise when fcc stacking sequences change.

\section*{II Effective 2 Dimensionality Arising from \\Stacking Faults,
hcp, 4H and 9R Structures}

First we will consider a simple case where a stacking fault is introduced in fcc stacking (ABC repeat) by replacing one of the trilayers ABC by a quintuple layer AABCC. The quintuple layer is a locally stable (metastable) state. Local stability comes from a direct H$_2$ molecule type bonding between s-orbitals of atoms in registry in AA bilayer. Similarly for CC bilayer. 

In a simple tight binding analysis we find that AA layer in isolation has a depleted density of states at the Fermi level with small electron and hole pockets. It effectively becomes a semimetal. Thus to a first approximation AA and CC bilayers isolate the middle B layer and makes it quasi 2 dimensional.

We expect presence  of a small density of such and similar stacking faults in Ag-Au NS. Such electronically isolated 2D triangular lattices  will become resorvoirs of singlet electron pairs, as we discussed before.

We have looked at hcp (AB repeat), 4H (ABCB repeat) and 9R (ABC.BCA.CAB repeat) from the above perspecive. In the cerorresponding stacking sequences we find layer units such as ABA (and permuted ones). We find possibility of next nearest neighbor AA binding, via interstitial orbitals of McAdon and Goddard \cite{Goddard}. This unusual bonding approximately decouples middle B layer and produces a corresponding cylindrical Fermi surface.

We find that the above type of decoupling can be done in different (energetically equivalent) ways. This allows \textit{possibility of dynamical decoupling or spontaneous symmetry breaking} and suggests rich new physics in this apparently simple system \cite{GBUnpublished}.

\end{document}